\documentclass[conference]{IEEEtran}

\ifCLASSINFOpdf
  \usepackage[pdftex]{graphicx}
\else
\fi

\usepackage{booktabs}
\usepackage{multirow}

\begin{document}
\title{On Cloud-Based Engineering of Dependable Systems}

\author{Sami~Alajrami \\ School~of~Computing~Science \\ Newcastle~University \\s.h.alajrami@ncl.ac.uk}


%Telephone: (800) 555--1212, Fax: (888) 555--1212}


\maketitle

%for forcing page numbers
%\thispagestyle{plain}
%\pagestyle{plain}

\begin{abstract}
The cloud computing paradigm is being adopted by many organizations in different application domains as it is cost effective and offers a virtually unlimited pool of resources. Engineering critical systems can benefit from clouds in attaining all dependability means: fault tolerance, fault prevention, fault removal and fault forecasting. Our research aims to investigate the potential of supporting engineering of dependable software systems with cloud computing and proposes an open, extensible, and elastic cloud-based software engineering workflow system which represents and executes software processes to improve collaboration, reliability and quality assurance, and automation in software projects.\\ 
\end{abstract}

\begin{IEEEkeywords}
Cloud Computing, Cloud Workflow Systems, Dependable Systems, Software Engineering.
\end{IEEEkeywords}

% \fi
\IEEEpeerreviewmaketitle

\section{Introduction}

Critical software systems (e.g., railway control and spacecraft navigation systems) have raised the need for dependable software. Dependability of a software system refers to many attributes such as: reliability, availability, safety and security. To produce a dependable software product, it is important to consider the different activities (and their artefacts) in the software development process followed. This adds to the complexity and cost of the software process. Depending on the trade-off between the cost of achieving dependability and the cost that might be incurred in case of a failure, some dependability attributes might be ignored leading to less dependable software.      

While some software engineering activities require minimal computing resources to be performed, other activities might demand extensive computing resources (processors, memory, storage, and networking). For example, editing a requirements document can be performed on an average desktop while a model checking a complex model might need a cluster of servers. Therefore, cloud computing has a potential to support dependable software development processes as it provides a large variance of scalable and cost-effective resources to meet the various demands of software process activities. Cloud would potentially reduce the cost of achieving dependability. 

Cloud providers offer three service models: a) Infrastructure as a Service (IaaS), b) Platform as a Service (PaaS), c) Software as a Service (SaaS). Depending on how much control on physical resources is attained, cloud deployment can be categorized into three models: a) Public cloud (services are provided to various clients)  , b) Private cloud (services are used by one organization), c) Hybrid cloud (a mix of the previous two). A typical workflow of critical software development process involves several activities with different execution requirements and constraints. The hybrid model can meet these requirements while still maintaining cost-efficiency. For example some activities involve using or generating confidential artefacts (models, designs, code, etc...), for such activities, using private cloud can satisfy the confidentiality requirement. On the other hand, for other activities -where confidentiality is not a requirement-, public cloud could be cheaper to use.    

Cloud has attracted many domains due to its cost-efficiency, scalability, and simplicity. However, using cloud for software engineering has not been in fancy and the clouds are still being seen as a deployment solution rather than a development platform. This is because none of the cloud's service models are ready for software engineering processes and activities~\cite{S-SEaaS}. 

This paper outlines a PhD study focusing on developing and implementing an architecture for cloud-based software engineering workflow system. The current state of art of using cloud for software engineering and initial experiments are briefly described in section II. The proposed architecture is represented in section III. The last section outlines the conclusion and future work.

\section{Background}

Some tool makers have started providing cloud-based software development tools. Examples include: Assembla\footnote{http://www.assembla.com}, Github\footnote{http://www.github.com}, Codenvy\footnote{http://www.codenvy.com}, IBM Jazz\footnote{http://www.jazz.net}. The common lack in all of these tools is that their focus is narrow to one or some of the software development process activities. Research has also been focusing on individual activities of software engineering. Testing is an example of activities that have been investigated. The testing performance of YETI -an automated random testing tool- has been improved after migrating it to the cloud~\cite{MF-yeti}. Cloud9~\cite{Bucur:Cloud9} is a parallel symbolic-execution-based testing platform on the cloud, which promises to make testing faster, cheaper, and more efficient. Authors in~\cite{M-SEaaSHPC} propose Software Engineering as a Service (SEaaS) framework for HPC. The framework offers different agile services to fit different project settings. However, this framework focuses more on project management aspects. The author of ~\cite{S-SEaaS} introduces the vision of a Software Engineering as a Service (EaaS) on a pay as you go model. 

The use of model-based development approaches for critical systems has been investigated in literature. Tool chains for automotive systems development activities have been proposed~\cite{my.erts2012},~\cite{herzner2007modelbased}. Similar work for aeronautics is presented in~\cite{aero.erts2012}. Engineering of modern critical systems that meet stringent dependability requirements needs the application of a comprehensive set of methods and tools, substantial computing (often shared) resources to support them as well as an involvement of large groups of developers and other stakeholders. The development artefacts need to be shared among developers, delivered to the customers and used for building the assurance cases. However, the current practice is not meeting all these needs. Provisioning of software development platforms in the cloud should proof beneficial to reduce upfront investments and software production costs, optimise performance of CASE tools using cloud's elasticity, and bridge the gap between development and deployment environments.
 
As part of this study, an initial experiment was conducted to better understand the challenges that software engineering would face in the cloud. e-Science central (e-SC)~\cite{HSPJ-eSC} -a science as a service workflow system- has been used to execute a software verification workflow on the cloud. e-SC was preferred to other workflow systems such as: Taverna~\footnote{http://www.taverna.org.uk}
because of its features such as: portability to different clouds, collaboration support, provenance, versioning, and exposure to developers. 

Model checkers are computing intensive software verification tools. The first experiment used the model checker Spin~\cite{G-spin} which has been integrated into the e-SC platform as a workflow block. Furthermore, a workflow has been created to receive a PROMELA model file and an instructions file as an input to perform the model checking on the cloud. When the workflow is executed, e-SC assigns it to a single workflow engine instance on which it will be executed.

In fact, due to the mechanism that e-SC uses for executing workflows, wrapping the model checker Spin into a workflow block was exactly the same as deploying it on a web server. The limitation of this approach is that it does not best utilize the cloud's elasticity as the resources will be limited to those of the single virtual machine executing a specific e-SC workflow. Therefore, executing parallel/distributed model checkers on multiple cloud instances has been considered.

The second experiment was to wrap a parallel model checker into an e-SC workflow block in order to investigate how it will benefit from the cloud elasticity. Parallel model checkers require heavy synchronization between all participating computing nodes. The parallel model checker Divine~\cite{BBCR10} was selected for this experiment as it is an open source model checker and it verifies models in multiple input formats such as: UPPAAL, LLVM, and DVE. In addition, Divine work on a cluster of multiple nodes which makes it suitable for utilizing cloud elasticity. 

Since e-SC hides the clouds from users (which is good for simplification), it was not possible to build an elastic workflow block that scale resources up and down as required using e-SC API. Hence, an e-SC elastic block has been developed to provision a specific number of customized  instances on Amazon EC2 (IaaS) on the fly. This elastic block has then been used to execute the parallel model checker Divine on a cluster of N nodes on EC2. Since it is hard to predict how much resources a model would require to be checked, it is possible that the allocated resources for a model checking task will not be sufficient. A set of execution parameters has been used in the elastic block to allow allocating more machines to the model checking task when a pre-configured timeout is expired without successfully completing it. These parameters are:

\begin{itemize}
\item Machine type: the Amazon EC2 machine type which specifies available resources (processing power, memory, storage, and network).
\item Number of Instances: the number of cloud instances to start the execution with.
\item Timeout: the initial execution time in hours.
\item Scaling type: to calculate the number of instances to be used when scaling up (either exponential or linear).
\end{itemize} 

The experiment has been performed using a model of the Peterson mutual exclusion algorithm provided as an example within the Divine package. Two different Amazon machine types have been used, also the number of threads used for model checking has been used as an internal Divine parameter. The results showed that adding more resources does not simply guarantee performance improvement. Therefore, special care should be taken to achieve performance gain from elasticity when deploying third-party ready-made software engineering tools.

The initial experiments have helped in identifying two types of challenges. First, those that face adopting e-SC and alike systems for software engineering (lack of user interaction, elasticity support, interoperability of tools, and using data flow approach). Second, general challenges related to utilizing cloud for software engineering such as:
\begin{itemize}
\item Implementation details of software development tools that might affect performance gain.
\item Organizations follow different processes, hence, flexibility in defining and customizing processes is required.
\item Software engineering activities are performed in practice using trial and error approach, hence, the tools  should be easy to run/rerun, adjust and rerun.
\item The cloud introduces an extra (hidden) level of complexity, hence, good diagnostics and reporting is essential. 
\end{itemize} 

\section{Proposed Architecture}  

Software processes are instances of business processes which involve activities that are either human interactions and decisions or computational tasks performed by software development tools. The complete set of activities involved in the process of producing a software product compensates for a software development workflow. Each activity is the responsibility of a role which is performed by an individual or a team. Based on that, an architecture for cloud-based software engineering workflow system is proposed.

The purpose of the workflow system is to support execution of software processes on the cloud. A software process workflow consists of several activities, in which each activity can be either: a) a manual task or decision point performed by a human, b) an automated task performed by a tool, c) a complete sub-workflow. To provide better controllability on the workflow execution, the granularity level is set to individual activities. i.e. each activity is executed independently from others provided that its preconditions are satisfied (e.g. an input is available).   

\begin{figure}[hc]
\begin{center}
\includegraphics[scale=0.55]{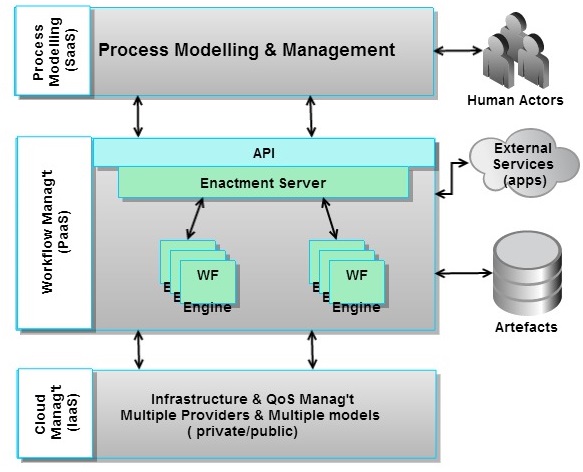}
\caption{Proposed architecture to support SE in the cloud\label{fig:architecture}}
\end{center}
\end{figure}

The architecture as illustrated in Figure~\ref{fig:architecture} is derived from the three service models of cloud. The cloud-based software engineering workflow system consists of three logical layers: 

\begin{itemize}
\item Software Process Modelling: The first step to use a workflow system for software engineering is to capture and model the software development process followed by an organization. Software process models such as the waterfall model are high level abstract representations of software processes and are tailored differently by different organizations. Therefore, it is not practical to provide a limited set of pre-configured models from which an organization can choose.  Alternatively, a complete software process modelling language is required to capture the main characteristics of the set of activities involved in the software process. Suitability of a process modelling language depends on how best the language features fit for the purpose of modelling. In the context of this architecture, the most important features of a modelling language are: a) expressiveness, to capture all the configurations of the software process (including cloud configurations), b) executability, in order to support process execution.

\item Workflow Management: This layer mimics the PaaS layer and it interacts with the process definition (SaaS) layer through an API so that it can work with any SaaS layer implementation (e.g. web/mobile apps). The API provides an interface to interact with the workflow enactment service. This service interprets the process model and instantiate it before allocating it to one or more workflow engines (deployed on different clouds) to be executed.
    
\item Cloud Management: Since the workflow system is deployed on the cloud, it has to address some cloud issues. The system need to be portable (i.e. can work on any cloud provider or private clouds). This can be achieved by using portable implementation techniques for the workflow management layer. Furthermore, smart process execution scheduling is needed to support optimization of cost and performance when distributing activities from one process into different clouds to be executed. Finally, efficient techniques are needed to control the QoS of the acquired cloud resources. 
\end{itemize}

\section{Conclusion \& Future work}

This paper briefly presents the potential of using clouds for engineering dependable systems, and outlines the initial architecture that is being developed. Next, a prototype for experimentation purposes will be implemented in order to evaluate this architecture and identify any unseen challenges. 

\bibliographystyle{IEEEtran}
\bibliography{reference}

\end{document}